\documentclass[preprint,review,12pt]{elsarticle}
\usepackage{extsizes}
\usepackage[version=3]{mhchem}
\usepackage[left=1.7cm, right=1.7cm, top=1.785cm, bottom=2.0cm]{geometry}
\usepackage{times,mathptmx}
\usepackage{gensymb}
\usepackage{sectsty}
\usepackage{epsfig}
\usepackage{lastpage}
\usepackage{float}
\usepackage{fnpos}
\usepackage[english]{babel}
\usepackage{array}
\usepackage{droidsans}
\usepackage{charter}
\usepackage[T1]{fontenc}
\usepackage[usenames,dvipsnames]{xcolor}
\usepackage{setspace}
\usepackage[compact]{titlesec}
\usepackage{array}
\usepackage{color}
\usepackage{textcomp}
\usepackage{amsmath}
\usepackage{amssymb}
\usepackage{amsthm}

\usepackage{epstopdf}

\makeatletter
\def\ps@pprintTitle{%
     \let\@oddhead\@empty
     \let\@evenhead\@empty
     \def\@oddfoot{}%
		 \let\@evenfoot\oddfoot}
\makeatother

\begin{document}

\begin{frontmatter}

\title{Study of network composition in interpenetrating polymer networks of poly(N isopropylacrylamide) microgels: \\ the role of poly(acrylic acid)}

\author[a,b]{Valentina Nigro \corref{corr1}}
\author[a,b]{Roberta Angelini\corref{corr1}}
\author[b]{Benedetta Rosi\fnref{Perugia}}
\author[d]{Monica Bertoldo \fnref{ISOF}}
\author[d]{Elena Buratti}
\author[e]{Stefano Casciardi}
\author[a,b]{Simona Sennato}
\author[a,b]{Barbara Ruzicka\corref{corr1}}

\address[a]{Istituto dei Sistemi Complessi del Consiglio Nazionale delle Ricerche (ISC-CNR), sede Sapienza, Pz.le Aldo Moro 5, I-00185 Roma, Italy} 

\address[b]{Dipartimento di Fisica, Sapienza Universit$\grave{a}$  di Roma, P.le Aldo Moro 5, 00185 Roma, Italy}
\address[d]{Istituto per i Processi Chimico-Fisici del Consiglio Nazionale delle Ricerche (IPCF-CNR), Area della Ricerca, Via G.Moruzzi 1, I-56124 Pisa, Italy}
\address[e]{Department of Occupational and Environmental Medicine, Epidemiology and Hygiene, National Institution for Insurance Against Accidents at Work (INAIL Research), Via Fontana Candida 1, Monte Porzio Catone, 00040 Rome, Italy}

\fntext[Perugia]{Present address: Dipartimento di Fisica e Geologia, Universit$\grave{a}$ di Perugia, Via A. Pascoli, 06123 Perugia (Italy)}
\fntext[ISOF]{Present address:Istituto per la Sintesi Organica e la Fotoreattivit$\grave{a}$ del Consiglio Nazionale delle Ricerche (ISOF-CNR), via P. Gobetti 101, 40129 Bologna, Italy}

\cortext[corr1]{Corresponding authors. E-mail addresses: valentina.nigro@uniroma1.it (V. Nigro); roberta.angelini@cnr.it (R. Angelini);  barbara.ruzicka@cnr.it (B. Ruzicka). Tel: +39 06 4991 23469}

\begin{abstract}
\textit{Hypothesis:} The peculiar swelling behaviour of poly(N-isopropylacrylamide) (PNIPAM)-based responsive microgels provides the possibility to tune both softness and volume fraction with temperature, making these systems of great interest for technological applications and theoretical implications. 
Their intriguing phase diagram can be even more complex if poly(acrylic acid) (PAAc)  is interpenetrated within PNIPAM network to form Interpenetrating Polymer Network (IPN) \footnote{IPN: Interpenetrating Polymer Network} microgels that exhibit an additional pH-sensitivity. The effect of the PAAc/PNIPAM polymeric ratio on both swelling capability and dynamics is still matter of investigation. \\
\textit{Experiments:} Here we investigate the role of PAAc in the behaviour of IPN microgels across the volume phase transition through dynamic light scattering (DLS) \footnote{DLS: Dynamic Light Scattering}, transmission electron microscopy (TEM) \footnote{TEM:transmission electron microscopy} and electrophoretic measurements as a function of microgel concentration and pH. \\ 
\textit{Findings:} Our results highlight that aggregation is favored at increasing weight concentration, PAAc content and pH and that a crossover PAAc content $C^*_{PAAc}$ \footnote{$C_{PAAc}$: weight content (\%) of the PAAc network within each IPN microgel} exists above which the ionic charges on the microgel become relevant. Moreover we show that the softness of IPN microgels can be tuned ad hoc by changing the PAAc/PNIPAM ratio. These findings provide new insights into the possibility to control experimentally aggregation properties, charge and softness of IPN microgels by varying PAAc content.\\

\end{abstract}


\end{frontmatter}

\section*{Introduction}
\label{Introduction}

The novel class of responsive microgels has recently become very popular since their smart responsivity to external stimuli makes them very attractive for industrial applications \cite{VinogradovCurrPharmDes2006, DasAnnRevMR2006, ParkBiomat2013, HamidiDrugDeliv2008, SmeetsPolymSci2013, SuBiomacro2008} and excellent model systems for exploring the exotic behaviours emerging in soft colloids due to their softness \cite{LikosJPCM2002,RamirezJPCM2009,HeyesSM2009}. Their interparticle potential and their effective volume fraction can be easily managed through unusual control parameters such as temperature, pH or solvent, allowing to explore unusual phase-behaviours \cite{WangChemPhys2014, PritiJCP2014, HellwegCPS2000, PaloliSM2013, LyonRevPC2012, WuPRL2003}, significantly far from those of conventional hard colloids \cite{PuseyNat1986, ImhofPRL1995, PhamScience2002, EckertPRL2002, LuNat2008, RoyallNatMat2008, RuzickaNatMat2011, AngeliniNC2014, BanchioJCP2008, GapinskiJCP2009}.
The deep investigation in the last years has shown how responsive microgels based on poly(N-isopropylacrylamide) (PNIPAM) undergo a reversible Volume Phase Transition (VPT)\footnote{VPT: Volume Phase Transition} at about 305 K that drives the system from a swollen hydrated state to a shrunken dehydrated one, as a consequence of the coil-to-globule transition of NIPAM chains \cite{MaColloidInt2010}. It has been shown that the driving force for swelling can be estimated from the properties of linear PNIPAM solutions, while the microgel elasticity opposing swelling is mainly due to the network topology dependent on the cross-linker concentration \cite{ShibayamaJCP1992, KratzBerBunsenges11998, KratzPolymer2001}. The typical swelling/shrinking behaviour of any PNIPAM-based microgel leads to intriguing phase diagrams which may be even more complex if other polymers, sensitive to different external stimuli, are copolymerized or interpenetrated to obtain multi-responsive microgels. In particular PNIPAM microgels containing poly(acrilic acid) (PAAc), have an additional pH-sensitivity that controls mutual polymer/polymer and polymer/solvent interactions.
In the case of PNIPAM-co-PAAc microgels\footnote{PNIPAM-co-PAAc: PNIPAM copolimerized with PAAc} the response strictly depends on the mutual interference between PNIPAM and PAAc \cite{KratzColloids2000, KratzBerBunsenges21998, JonesMacromol2000, XiongColloidSurf2011, MengPhysChem2007, LyonJPCB2004, HolmqvistPRL2012, DebordJPCB2003}.
On the contrary interpenetration of the hydrophilic PAAc and the homopolymeric PNIPAM networks (IPN PNIPAM-PAAc microgel) \cite{HuAdvMater2004, XiaLangmuir2004, XiaJCRel2005, ZhouBio2008, XingCollPolym2010, LiuPolymers2012, NigroJNCS2015, NigroJCP2015, NigroCSA2017, NigroSM2017}, provides independent sensitivity to temperature and pH, retaining the same Volume Phase Transition Temperature (VPTT) \footnote{VPTT: Volume phase Transition Temperature} of pure PNIPAM microgel and allowing to make the two networks more or less dependent by changing pH. Interestingly, IPN microgels allow to control their elastic properties by changing the solution pH, the polymeric ratio PNIPAM/PAAc and the cross-linking degree of any polymeric network. 
Notwithstanding the IPN microgel potentialities, knowledge of their behaviour from a fundamental point of view is still very limited. Hu et al. reported a comparison between the hydrodynamic and giration radii in very dilute conditions \cite{XiaLangmuir2004} and investigated the phase behaviour and viscosity for controlled drug release as a function of temperature \cite{HuAdvMater2004, XiaJCRel2005}. 
The viscoelastic behaviour has been studied by Zhou et al. \cite{ZhouBio2008} who tested also the in vivo controlled release. The role of softness in the glass formation has been assessed by 
Mattsson et al. \cite{MattssonNature2009} who showed how these deformable colloidal particles can exhibit the same variation in fragility as that observed in molecular liquids. 
Liu et al. \cite{LiuPolymers2012} performed a systematic morphological study on IPN particles in highly dilute conditions. Particle synthesis was controlled through different techniques and the hydrodynamic radius was obtained as a function of temperature for different pH and PAAc contents.
Moreover IPN microgels have been investigated by our group through different techniques: the structural relaxation and the local structure of low concentration IPN samples at pH 5 and pH 7 have been respectively reported in Ref. \cite{NigroJNCS2015} and Ref. \cite{NigroJCP2015}. The dependence on different solvents (H$_2$O and D$_2$O), to explore the role of H-bonds, has been reported in Ref. \cite{NigroSM2017} at different length scales. Finally in Ref. \cite{NigroCSA2017} the experimental radius of PNIPAM and IPN at fixed PAAc concentration has been compared with the one expected from the Flory-Rehner theory.

At variance with previous studies \cite{XiaLangmuir2004, HuAdvMater2004, XiaJCRel2005, ZhouBio2008,MattssonNature2009,LiuPolymers2012, NigroJNCS2015, NigroJCP2015, NigroSM2017, NigroCSA2017} the present research aims to understand the role played  by poly(acrylic acid) on the relaxation dynamics and on the aggregation process across the VPT.
Despite the work done up to now on IPN microgels, to the best of our knowledge a similar study has never been reported before.
We demonstrate that with increasing PAAc content an increase of charge density favours the formation of aggregates due to the interplay between attractive and repulsive interactions that can be triggered by changing both PAAc and pH. Moreover a critical PAAc concentration is found that signs the existence of two different behaviours, this phenomenology can be ascribed to the increasing relevance of the ionic charge.

In the following we first characterize the microgel particles through dynamic light scattering and transmission electron microscopy measurements to gain information on hydrodynamic radii and particles morphology. A comparison with a Flory-Rehner model, which well reproduce the temperature behaviour of the measured radii, allows to have a qualitative insight on the number of counterions per polymer chain that increases with increasing PAAc. Then we find a huge growth of the relaxation time above the VPTT at high PAAc values and pH, clear evidence of the formation of large aggregates. Through mobility measurements we can relate this behaviour to the observed increase of mobilities towards more negative values, demonstrating the role played by PAAc and charge density on aggregation. Finally we prove that it is possible to tune ad hoc particle softness by tuning PAAc.

\section*{Experimental Methods}
\label{Experimental Methods}

\subsection*{Sample preparation}

\paragraph*{Materials}
N-isopropylacrylamide (NIPAM) \footnote{NIPAM: N-isopropylacrylamide} (Sigma-Aldrich), purity 97 \%, and N,N'-methylene-bis-acrylamide (BIS)\footnote{BIS: N,N'-methylene-bis-acrylamide} (Eastman Kodak), electrophoresis grade, were purified by recrystallization from hexane and methanol, respectively, dried under reduced pressure (0.01 mmHg) at room temperature and stored at 253 K. Acrylic acid (AAc)\footnote{AAc: Acrylic acid} (Sigma-Aldrich), purity 99 \%, with 180-220 ppm of MEHQ as inhibitor, was purified by distillation (40 mmHg, 337 K) under nitrogen atmosphere on hydroquinone and stored at 253 K. Sodium dodecyl sulfate (SDS)\footnote{SDS: Sodium dodecyl sulfate }, purity 98 \%, potassium persulfate (KPS)\footnote{KPS: potassium persulfate}, purity $\geq$ 98 \%, ammonium persulfate (APS)\footnote{APS: ammonium persulfate }, purity 98 \%, N,N,N',N'-tetramethylethylenediamine (TEMED), purity 99 \%, ethylenediaminetetraacetic acid (EDTA)\footnote{EDTA: ethylenediaminetetraacetic acid}, purity $\geq$ 98.5 \%, and NaHCO$_{3}$\footnote{NaHCO$_{3}$: sodium hydrogen carbonate}, purity 99.7-100.3 \%, were all purchased from Sigma-Aldrich and used as received. Ultrapure water (resistivity: 18.2 M$\Omega$$\cdot$cm at 298 K) was obtained with Millipore Direct-Q\textsuperscript{\textregistered} 3 UV purification system. All other solvents (Sigma Aldrich RP grade) were used as received. Dialysis tubing cellulose membrane (Sigma-Aldrich), cut-off 14,000 Da, was washed in running distilled water for 3 h, treated at 343 K into a solution at 3.0 \% weight concentration of NaHCO$_{3}$ and 0.4 \% of EDTA for 10 min, rinsed in distilled water at 343 K for 10 min and finally in fresh distilled water at room temperature for 2 h.

\paragraph*{Synthesis of PNIPAM and IPN microgels}
PNIPAM microgels were synthesized by precipitation polymerization with (24.162 $\pm$ 0.001) g of NIPAM, (0.4480 $\pm$ 0.0001) g of BIS and (3.5190 $\pm$ 0.0001) g of SDS, solubilized in 1560 mL of ultrapure water and transferred into a 2000 mL four-necked jacked reactor equipped with condenser and mechanical stirrer. The solution was deoxygenated by bubbling nitrogen for 1 h and heated at (343 $\pm$ 1) K. (1.0376 $\pm$ 0.0001) g of KPS (dissolved in 20 mL of deoxygenated water) was added to initiate the polymerization and the reaction was allowed to proceed for 16 h. The resultant PNIPAM microgel was purified by dialysis against distilled water with frequent water change for 2 weeks. In the second step IPN microgels were synthesized by a sequential free radical polymerization method \cite{XiaLangmuir2004} with (140.08 $\pm$ 0.01) g of the PNIPAM dispersion at the final weight concentration of 1.06 \%. 5 mL of AAc and (1.1080 $\pm$ 0.0001) g of BIS were added into the preformed PNIPAM microparticles in the temperature range where PNIPAM particles are swollen (T=294 K), allowing the growth of the PAAc network inside them. The mixture was diluted with ultrapure water up to a volume of 1260 mL and transferred into a 2000 mL four-necked jacketed reactor kept at 294 K by circulating water and deoxygenated by bubbling nitrogen inside for 1 h. 0.56 mL of TEMED were added and the polymerization was started with (0.4447 $\pm$ 0.0001) g of ammonium persulfate. Three different samples were prepared at three PAAc/PNIPAM ratio composition by stopping the reaction at the suitable degree of conversion of AAc. The samples were purified by dialysis against distilled water with frequent water changes for 2 weeks, iced and  liophilized up to 1 \% weight concentration. The synthesized particles, namely PNIPAM and IPN microgels, were analysed by ATR FT-IR and $^1$H-NMR spectroscopies, as well as by elemental analysis to assess their chemical composition and the exact PAAc content \cite{VillariCPC2018}. The three investigated samples have the following PAAc weight concentrations ($C_{PAAc}$): $C_{PAAc}=2.6 \%$, $C_{PAAc}=10.6 \%$, $C_{PAAc}=19.2 \%$ \cite{NotaCampioni}.

Their polydispersity is found around 10-15\% for PNIPAM and 15-20\% for IPN microgels. PNIPAM polydispersity values are due to the SDS concentration used during synthesis C$_SDS$ =2.25 g/L =7.81 mM in order to obtain nanosized particles. In fact while monodisperse microsized particles are obtained by using a  C$_SDS$ far below the critical micelle concentration (CMC=2.36 g/L =8.18 mM) C$_SDS$ $\ll$ CMC, nanosized particles, as in the present work, are synthesized when the amount of SDS approaches the critical micelle concentration.  These samples are more homogeneously structured than the samples prepared in low surfactant concentration but with an increased polydispersity \cite{AnderssonJPolymSci2006}. In the case of IPN, particle size distribution is further enlarged if merging of PNIPAM particles is not fully suppressed during the polyacrilic acid polymerization \cite{VillariCPC2018}. Samples at different weight concentrations (\%), in the following referred as $C_w$, were obtained by diluting in H$_2$O  the sample at 1\% concentration. Samples at pH 3.5 and pH 7.5 were obtained by the addition of HCl \footnote{HCl: Hydrochloric acid} or NaOH \footnote{NaOH: Sodium hydroxide}, respectively, to the samples at pH 5.5 obtained from the synthesis.

\subsection*{Transmission Electron Microscopy}
\label{TEM}

Transmission Electron Microscopy (TEM) characterization was employed to study PNIPAM and IPN microgels morphology. All the samples for TEM measurements have been prepared by deposition at room temperature of 20 $\mu$L of microgel suspensions diluted up to $C_w$=0.1 \% in MilliQ water, on a 300-mesh copper grid for electron microscopy covered by thin amorphous carbon film.  Immediately after deposition, the excess of liquid was removed by touching the grid with a piece of filter paper. Samples were dried for 5 minutes before staining by addition of 10 $\mu$L  of  2 \% aqueous phosphotungstic acid (PTA)(pH-adjusted to 7.3 using 1 M NaOH).   Measurements were carried out by using a FEI TECNAI 12 G2 Twin (FEI Company, Hillsboro, OR, USA), operating at 120 kV and equipped with an electron energy filter (Gatan image filter) and a slow-scan charge-coupled device camera (Gatan multiscan).
Statistical analysis of TEM images to determine the average diameter and particle size distribution were performed by Image J software by measuring the cross-sectional area of the particles and convert them to an equivalent spherical diameter. A minimum of 100 particles collected on  different captured images with the same magnification has been considered. The average size has been determined by considering the mean value obtained by a gaussian fit on the particle size distribution, the reported error being the statistical error of the mean.

 \subsection*{Dynamic Light Scattering}
 \label{Dynamic Light Scattering}

\begin{figure*}[t]
\centering
\includegraphics[width=13cm]{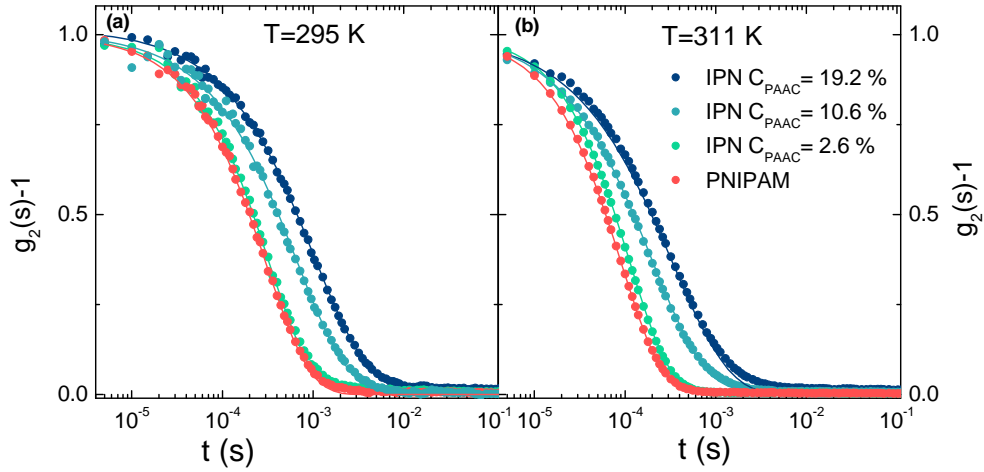}
\caption{\label{fig:DLS}Normalized intensity autocorrelation
functions for PNIPAM and IPN microgels at different PAAc contents at pH 5.5, at $C_w=0.1$ \% and
$\theta$=90\textdegree, corresponding to Q=0.018 nm$^{-1}$, at temperature (a) below and (b) above the VPTT. Solid lines are fits according to Eq.(\ref{Eqfit}).}
\end{figure*}

Multiangle Dynamic Light Scattering (DLS) measurements have been performed using an optical setup based on a monochromatic and polarized beam emitted from a solid state laser (100 mW at $\lambda$=642 nm) and focused on the sample in a cylindrical VAT for index matching and temperature control. The scattered intensity is collected by five single mode optical fibers at five different scattering angles, namely $\theta$=30\textdegree, 50\textdegree, 70\textdegree,
90\textdegree, 110\textdegree, corresponding to five
scattering vectors $Q$ in the range (0.0067 nm$^{-1}$ $\leq$ Q $\leq$ 0.021 nm$^{-1}$), according to the relation Q=(4$\pi$n/$\lambda$) sin($\theta$/2).
The normalized intensity autocorrelation functions $g_2(Q,t)=<I(Q,t)I(Q,0)>/<I(Q,0)>^{2}$ are obtained with a high coherence factor close to the ideal unit value.
The experiment has been performed on aqueous suspensions of PNIPAM and IPN microgels at three PAAc contents ($C_{PAAc}$=2.6 \%, $C_{PAAc}$=10.6 \%, $C_{PAAc}$=19.2 \%) in the temperature range T=(293$\div$313) K across the VPT, at four weight concentrations ($C_w$=0.1 \%, $C_w$=0.3 \%, $C_w$=0.5 \% and $C_w$=0.8 \%) and three different pH values (pH 3.5, 5.5 and 7.5). The reported data have been obtained by averaging five repeated set of measurements. Particle sizes have been determined  from the decay constant
$\Gamma(Q)=D q^2$  obtained through the analysis of the intensity correlation functions.
The behaviour of the normalized intensity autocorrelation functions collected at $\theta$=90\textdegree~ for PNIPAM and IPN microgels at different PAAc contents and at low weight concentration ($C_w$=0.1 \%) is reported in Fig.\ref{fig:DLS}, below and above the VPTT. 
As commonly known, the intensity correlation function of most
colloidal systems is well described by the Kohlrausch-Williams-Watts expression \cite{KohlrauschAnnPhys1854, WilliamsFaradayTrans1970}:

\begin{equation}
g_2(Q,t)=1+b[(e^{-t/\tau})^{\beta}]^{2} \label{Eqfit}
\end{equation}

where $b$ is the coherence factor, $\tau$ \footnote{$\tau$: structural relaxation time} is the structural relaxation time and $\beta$ \footnote{$\beta$: stretching parameter} describes the deviation from the simple exponential decay ($\beta$ = 1) usually found in monodisperse systems. Indeed the distribution of the relaxation times in disordered materials leads to a stretching of the correlation functions characterized by an exponent $\beta$ $<$ 1 which can be related to the distribution of the relaxation times due to the polydispersity of the samples.

\subsection*{Electrophoretic Measurements}
\label{}

Electrophoretic mobility of microgel suspensions was measured by means of a MALVERN
NanoZetasizer apparatus equipped with a 5 mW HeNe laser (Malvern Instruments LTD, UK).
This instrument employs  traditional Laser Doppler Velocimetry  (LDV) implemented with Phase Analysis Light Scattering (PALS) for a more sensitive detection of the Doppler shift \cite{TscharnuterAppOpt2001}. LVD measurements are performed using the patented "mixed mode" measurement M3 where both a fast field (FF) and a slow field (SF) are applied. In FFR the field is reversed 25-50 times per second, thus making electro-osmosis insignificant and providing accurate mean mobility value. The SFR contributes extra resolution for a better distribution analysis \cite {MinorJCIS1997, ConnahJDST2002}.
The frequency shift  $\Delta \nu$  due to the mobility $\mu$ of the scattered particles under the action of the applied field E is measured by comparing the phase $\Phi$ of the scattered signal to that of a reference one, since $\Phi = \nu \cdot$ time.
The mobility $\mu= V/E$ \footnote{$\mu$: electrophoretic mobility} is then calculated from the relation $\Delta \nu=  2 V \sin (\theta /2)/\lambda)$ with $V$  the particle velocity, $ \theta $ the scattering angle and $\lambda$ the laser wavelength. By a preliminary conductivity measurement, the instrument establishes a suitable electric field for a good mobility detection.
Both PNIPAM and IPN samples at the different PAAc contents have been measured at $C_w$=0.05 \% and pH around 5.5.  Measurements have been performed by using the dedicated U-cuvette DTS1070, in a thermostated cell by performing  a ramp from 293 to 316 K with temperature step of 1 K and a thermalization time of 300 s at each step. Data presented here correspond to the mean values of the electrophoretic mobility distribution and are obtained by averaging three repeated set of measurements.

\section*{Results and Discussions}
\label{Results}

\subsection*{Sample characterization}
\paragraph*{Hydrodynamic radius}

\begin{figure}[t]
\centering
\includegraphics[height=7cm]{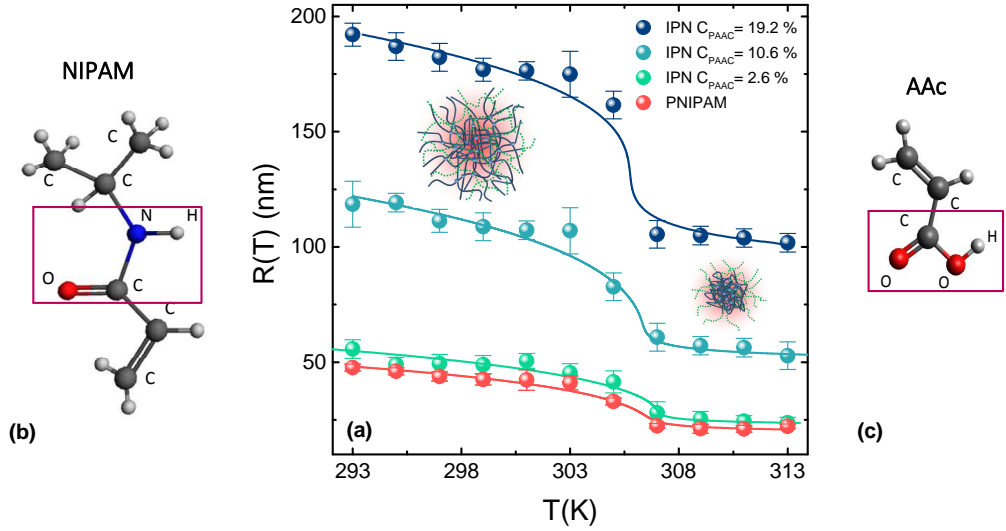}
\caption{(a) Hydrodynamic radii from DLS measurements for PNIPAM and IPN microgels at the indicated PAAc contents, pH 5.5 and $C_w=0.1$ \%. Solid lines are the best fits according to the Flory-Rehner model of Eq.(\ref{EqStateIonic}).(b) and (c) Sketch of NIPAM and AAc molecular structure respectively.}
\label{fig:RvsT}
\end{figure}

The temperature behaviour of the hydrodynamic radii of IPN and pure PNIPAM microgels has been obtained trough DLS at low weight concentration ($C_w=0.1$ \%) and is reported in Fig.\ref{fig:RvsT}. The typical swelling behaviour of microgel particles from a swollen to a shrunken state across the VPT is found and it is highly affected by the acrylic acid content, in very good agreement with what reported in Ref. \cite{LiuPolymers2012}. The increase of acrylic acid content determines sharper transitions. Notwithstanding some limits \cite{LopezSM2017} the most simple way to describe the microgels swelling is given by the Flory-Rehner theory \cite{Flory1953, FernandezBook2011} in terms of the total osmotic pressure $\pi$ \footnote{$\pi$: total osmotic pressure} inside the gel, consisting of a mixing contribution $\pi_m$ \footnote{$\pi_mi$: mixing contribution to the osmotic pressure} and an elastic component $\pi_e$ \footnote{$\pi_e$: elastic contribution to the osmotic pressure}. For ionic microgels, as IPN, additional contributions arising from the screened repulsion between polymer chains and from the osmotic pressure due to counterions confined inside the network, show up. However for charge density smaller than the value at which counterion condensation may take place the first of the two effects can be neglected \cite{QuesadaPerezSM2011}. Therefore for IPN microgels also the ionic contribution $\pi_i$ \footnote{$\pi_i$: ionic contribution to the osmotic pressure} to the osmotic pressure has to be taken into account. 
At the equilibrium condition the total osmotic pressure is $\pi=\pi_m+\pi_e+\pi_i$=0 and the swelling of a weakly ionized system can be theoretically explained through the equation of state:

\begin{equation}
ln(1-\phi)+\phi+\chi \phi^2+\frac{\phi_0}{N}[(\frac{\phi}{\phi_0})^{1/3}-(\frac{1}{2}+f)\frac{\phi}{\phi_0}]=0
\label{EqStateIonic}
\end{equation}

\noindent where $\phi_0$ \footnote{$\phi_0$: polymer volume fraction in the reference state} is the polymer volume fraction in the reference state, typically taken as the shrunken one \cite{LopezLeonPRE2007}, $\phi$ \footnote{$\phi$: polymer volume fraction within the particle} is the polymer volume fraction within the particle, which for isotropic swelling is related to the hydrodynamic radius through the relation $\frac{\phi}{\phi_0}=(\frac{R_0}{R})^3$ (with R$_0$ \footnote{$R_0$: hydrodynamic radius in the shrunken state} the hydrodynamic radius in the shrunken state) \cite{LopezLeonPRE2007, NigroCSA2017}, $N$ \footnote{N: number of segments occupied by a polymer chain between two cross-links} is the number of segments occupied by a polymer chain between two cross-links, $f$ \footnote{f: number of counterions per polymer chain} is the number of counterions per polymer chain, which is found to increase with PAAc content and $\chi$ \footnote{$\chi$: Flory polymer-solvent interaction parameter} is the Flory polymer-solvent interaction parameter, which has to be interpreted as an effective mean parameter accounting for polymer/solvent interactions, polymer/polymer interactions within each network and polymer/polymer interactions between different networks. It can be written as a power series expansion

\begin{equation}
\chi=\chi_1 (T)+\chi_2 \phi+\chi_3 \phi^2+\chi_4 \phi^3+\cdots
\label{Chi3}
\end{equation}

where $\chi_1$ \footnote{$\chi_1$: Flory parameter} is the Flory parameter, defined as $\chi_1=\frac{1}{2}-A(1-\frac{\theta}{T})$, $A$ is the parameter that provides rough details on the solvent quality, its increase indicates a decreasing goodness of the solvent as discussed in the Supplementary Material, $\theta$ \footnote{$\theta$: volume phase transition temperature} is the VPT temperature and $\chi_i$ (i=1,2,3,$\ldots$) \footnote{$\chi_i$: temperature independent coefficients} are temperature independent coefficients. The second-order approximation of Eq.(\ref{Chi3}) well describes the VPT of our microgel as previously reported for pure PNIPAM \cite{ErmanMacromol1986, LopezLeonPRE2007} and IPN microgels at acidic pH \cite{NigroCSA2017}. The Flory-Rehner model well reproduces the experimental data (symbols) and the goodness of the fits (lines) is evident from Fig.\ref{fig:RvsT}, the fit parameters are reported in Table\ref{tab:FRfit}.
The model gives values for the VPT temperature in agreement with those expected \cite{PeltonAdvColloid2000, NigroJNCS2015, NigroSM2017}. The number of ionized groups per chain f increases with increasing PAAc content indicating an increase of the charges of the microgel. In particular the f values suggest that for PNIPAM and for low (2.6 \%) PAAc content the ionic contribution can be neglected while it becomes more relevant at intermediate (10.6 \%) and high (19.2 \%) PAAc content where the swelling behaviour deviates from that of pure PNIPAM microgel. This deviation can be explained accounting for the topological constraints due to the networks interpenetration and to the ionic contribution to the osmotic pressure. Interestingly the A parameter increases with PAAc content indicating that mixing between polymer and solvent is not favored and swelling is inhibited by PAAc. These results strongly indicate that interpenetration of PAAc network affects the interaction between PNIPAM and water molecules, these polymer/solvent interactions can be therefore controlled through pH \cite{NigroCSA2017} or by varying PAAc content, as here highlighted for the first time. However it is worth noting that there is a discrepancy between the value of $\phi_0$ obtained with the Flory Rehner theory and other methods as deeply discussed in Ref. \cite{LopezSM2017}.

\begin{table}[t]
\small
  \begin{center}
     \begin{tabular}{c || c c c c c c c}
       \hline
	      {} & {$A$} & {$\phi_0$} & {$\theta (K)$} & {$\chi_2$} & {$\chi_3$} & {$N$} & {$f$}\\ 
				\hline
				\hline
        PNIPAM & -11.1 $\pm$ 0.9 & 0.70 $\pm$ 0.02 & 306.4 $\pm$ 0.6 & 0.19 $\pm$ 0.05 &  0.57 $\pm$ 0.06 & 118 $\pm$ 10 & 0.013 $\pm$0.005  \\
        IPN 2.6\%  & -10.04 $\pm$ 0.9 & 0.62 $\pm 0.02 $ & 306.5 $\pm$ 0.5 & 0.09 $\pm$ 0.04 & 0.76 $\pm$ 0.06 & 120 $\pm$ 11 & 0.15 $\pm$ 0.05 \\
        IPN 10.6\%  & -9.8 $\pm$ 0.8 & 0.74 $\pm$ 0.02 & 305.5 $\pm$ 0.5 & 0.12 $\pm$ 0.04 & 0.72 $\pm$ 0.05 & 116 $\pm$ 10 & 0.8 $\pm$ 0.2 \\
        IPN 19.2\%  & -2.1 $\pm$ 0.1 & 0.79 $\pm$ 0.02 & 305.5 $\pm$ 0.4 & 0.43 $\pm$ 0.01 & 0.25 $\pm$ 0.01 & 151 $\pm$ 9 & 2.7 $\pm$ 0.1\\ 
       \hline
	
     \end{tabular}
     \caption{Values of the parameter \textit{A}, the polymer volume fraction in the reference state $\phi_0$, the Flory temperature $\theta$, the number of lattice sites occupied by a polymer chain between two cross-links  $N$, the parameters $\chi_1$ and $\chi_2$ and the number of counterions per polymer chain $f$, as obtained from the best fit through Eq.(\ref{EqStateIonic}) of the hydrodynamic radii of Fig.\ref{fig:RvsT}.} \label{tab:FRfit}
   \end{center}
	\end{table}

\paragraph*{Morphology}
\begin{figure}[t]
\centering
\includegraphics[width=14cm]{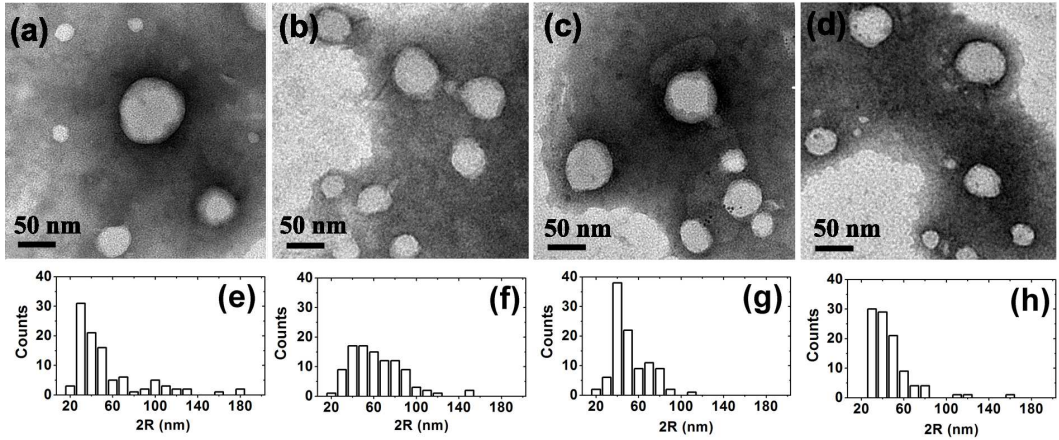}
\caption{\label{fig:TEM}TEM images obtained by negative PTA staining for (a) PNIPAM and IPN microgels prepared at three PAAc contents (b) $C_{PAAc}$=2.6 \%, (c) $C_{PAAc}$=10.6 \%, (d) $C_{PAAc}$=19.2 \%. Particle size distributions calculated on several collected TEM images for (e) PNIPAM and IPN microgels at three PAAc contents (f) $C_{PAAc}$=2.6 \%, (g) $C_{PAAc}$=10.6 \%, (h) $C_{PAAc}$=19.2 \%.}
\end{figure}

TEM images obtained by PTA staining for PNIPAM and IPN microgels prepared at the different PAAc contents ($C_{PAAc}$=2.6 \%, $C_{PAAc}$=10.6 \%, $C_{PAAc}$=19.2 \%) are shown in Fig.\ref{fig:TEM}, panel a-b-c-d, respectively. In all samples, several microgel particles appear as clear objects  with size ranging from 20 to 100 nm.  No significant differences between the morphology of IPN microgels with different PAAc contents appears from TEM images. In some cases, a sharp negative staining arises due to PTA accumulation close to microgel which creates a black halo around the particles. At a deeper observation of those microgels,  a thin dark grey region can be distinguished from the light grey inner part of the particles. This different appearence is probably caused by the PTA penetration over a small distance within the particle.  These results may indicate the presence of a more compact microgel inner core surrounded by a loose shell which is more permeable to PTA molecules both for PNIPAM and IPN microgels, probably due to less cross-linked chains with more dangling ends, as shown in different studies for PNIPAM-based microgels \cite{StiegerJCP2004, MasonPRE2005, ReuferEPJ2009,LedesmaCSA2015, RomeoSM2013}. A statistical analysis performed on TEM images is reported in panels e-f-g-h, for PNIPAM and IPN with $C_{PAAc}$=2.6 \%, $C_{PAAc}$=10.6 \% and $C_{PAAc}$=19.2 \%, respectively. The mean diameters from a statiscal analysis on the whole particle distribution are obtained as an average value with standard deviation: PNIPAM ($ 50 \pm 30$) nm, IPN 2.6 \% ($60 \pm 20$) nm,  IPN 10.6 \% ($40 \pm 20$) nm and IPN 19.2 \% ($ 50 \pm 20 $) nm. The wide particle distributions could be connected to flattening and deformation of samples on the support, as often occurs for soft materials, or to the intrinsic polydispersity of the samples that is related to the used SDS concentration during the synthesis as described in the Sample preparation paragraph.

\subsection*{Temperature behaviour}

\begin{figure}[t]
\centering
\includegraphics[height=7cm]{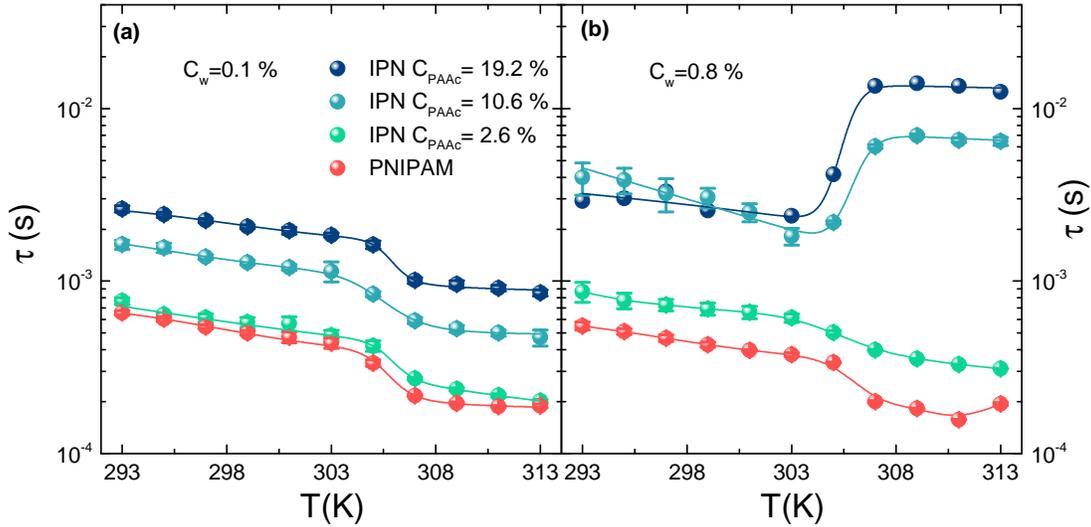}
\caption{Temperature behaviour of the relaxation time for PNIPAM and IPN microgels at the indicated PAAc contents, at (a) low ($C_w$=0.1 \%) and (b) high ($C_w$=0.8 \%) weight concentrations and pH 5.5. Solid lines are guides to eyes.}
\label{fig:tauPAAc}
\end{figure}

In order to understand the role played by the poly(acrylic acid) on the relaxation dynamics and on the aggregation process across the VPT, the temperature behaviour of the relaxation time, reported in Fig.\ref{fig:tauPAAc}, has been obtained through DLS by fitting the $g_2(Q,t)$ with Eq.(\ref{Eqfit}).
In the low weight concentration range (Fig.\ref{fig:tauPAAc}(a)) the well known dynamical transition associated to the VPT is evidenced for both PNIPAM and IPN microgels \cite{NigroJNCS2015, NigroCSA2017}: as temperature increases the relaxation time $\tau (T)$ slightly decreases up to the volume phase transition temperature above which it decreases to its lowest value, corresponding to the shrunken state indicating a fastening of the dynamics related to the reduced size of the particles and an increased diffusivity.
Surprisingly as the weight concentration increases (Fig.\ref{fig:tauPAAc}(b)) relaxation time above the VPTT is strongly affected by PAAc content: while at low PAAc it resembles that of pure PNIPAM, at higher PAAc it suddenly increases with temperature, thus indicating the formation of aggregates accompanied by an evident viscosity increase observed by eyes. 
This behaviour can be explained considering that above the VPTT, due to the reduced particle size, Van der Waals attraction becomes stronger, thus affecting the microgel aggregation. Moreover if particles are charged, as in the case of IPN microgels, also electrostatic interactions have to be taken into account, they are much stronger greater  the PAAc content is, as also evidenced by the increasing values of the $\it f$ parameter of table \ref{tab:FRfit} associated to the ionic contribution. In addition the collapse of NIPAM networks with temperature is supposed to favor the exposure of PAAc dangling chains and the formation of aggregates. 
\begin{figure}[!t]
\centering
\includegraphics[width=7cm]{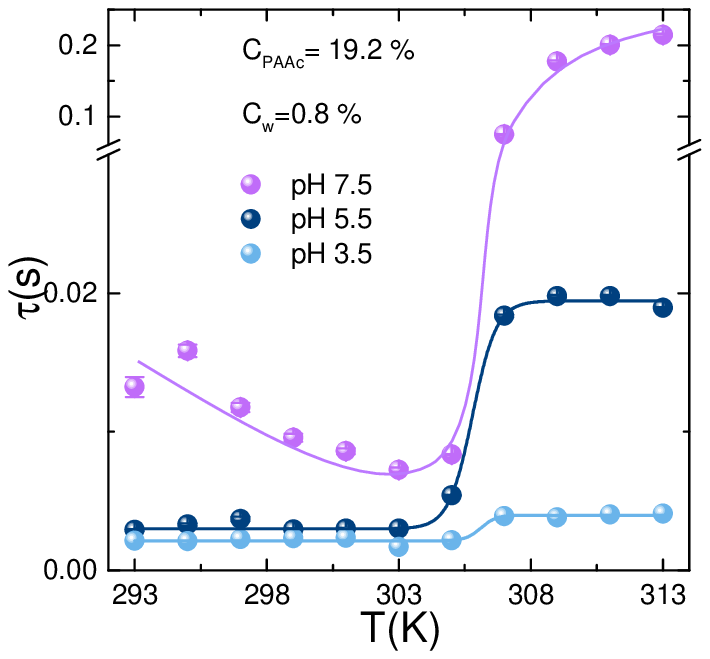}
\caption{Temperature behaviour of the relaxation time for IPN microgels at high weight concentration C$_w$=0.8 \% and high PAAc content (C$_{PAAc}$=19.2 \%), at pH 3.5, pH 5.5 and pH 7.5. Solid lines are guides to eyes.}
\label{fig:taupH}
\end{figure}

To test the importance of the charge in the aggregation process we performed measurements at three different pH since the overall charge can be tuned with pH due to the progressive deprotonation of the ionizable groups.
The temperature behaviour of the relaxation time for IPN microgels at C$_{PAAc}$=19.2 \%, $C_w$=0.8 \% and at pH 3.5, pH 5.5 and pH 7.5 is reported in Fig.\ref{fig:taupH}.  
At pH 7.5, the highest investigated pH, a huge growth of $\tau(T)$ above the VPTT is evident, indicating the formation of large aggregates. Indeed at this pH, the carboxylic groups COOH \footnote{COOH: carboxyl group} of PAAc (Fig.\ref{fig:RvsT}(c)) are dissociated into COO$^-$ and H-bonding between COOH  groups of AAc moieties belonging to different particles are not favored. As a consequence, the aggregation can be mainly ascribed to like-charge attraction since, at this pH, IPN behave as polyelectrolyte microgels where attraction can be interpreted as a result of counterion fluctuation due to the formation of temporary dipoles \cite{GrohnMacromol2000}. At intermediate pH 5.5 there is a fraction of COOH groups and a fraction, not negligible, of COO$^-$ groups, therefore aggregation can be described as a combination of both like-charge attraction and H-bonding interaction between COOH groups. Finally at pH 3.5 the COOH groups of PAAc are fully protonated (neutralized), electrostatic interactions are excluded and H-bondings with the amidic (CONH)\footnote{CONH: amidic group} groups of PNIPAM (Fig.\ref{fig:RvsT}(b)) inside the particles are largely favored. Nevertheless small aggregates above the VPTT are formed, suggesting that inter-particle interactions at high C$_{PAAc}$ are not excluded and can be mainly ascribed to strong H-bonding and hydrophobic interactions.

These results indicate that PAAc represents a good experimental control parameter to tune inter-particle interactions and aggregation: higher is the amount of acrylic acid interpenetrating the PNIPAM network, higher is the formation of aggregates and the viscosity increase.
Our findings immediately suggest the importance of the charge density.

\begin{figure}[!t]
\centering
\includegraphics[height=7cm]{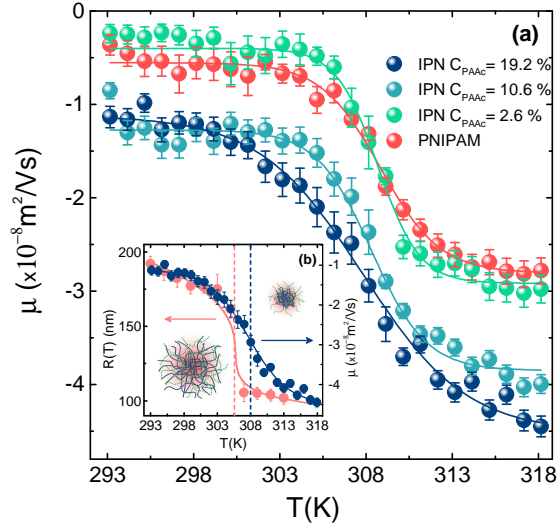}
\caption{(a) Electrophoretic mobility as a function of temperature for PNIPAM and IPN microgels at the indicated PAAc content at low weight concentrations and pH 5.5. Full lines are fits through the sigmoidal function reported in the text. (b) Comparison between electrophoretic mobility and hydrodynamic radius for IPN microgel at C$_{PAAc}$=19.2 \%. Vertical dashed lines represent the VPT temperature $\approx$ 305 K and the electrophoretic transition temerature T$_0\approx$ 308 K.}
\label{fig:mobility}
\end{figure}

To complement these observations and to gain information on charge density, we have performed electrophoretic measurements on PNIPAM and IPN microgels as a function of temperature (Fig.\ref{fig:mobility}). The mobility $\mu$ is affected by the volume phase transition and decreases as the suspension crosses the VPTT.  
For both PNIPAM and IPN microgels a sharp drop is observed around T$_0$ $\approx$ 308 K, as determined by a sigmoidal fit through the equation $y = A_2 + (A_1-A_2)/(1 + exp((T-T_0)/dT))$ whose fit parameters are reported in Table \ref{tab:MobFit}. For pure PNIPAM microgels the very low mobility below the VPTT reflects the low charge density of the particles, whereas above the VPTT it increases toward more negative values. It is in fact not surprising that a negative charge appears for particles obtained with the ionic initiator KPS. Considering that the negative electrical charges brought by the anionic sulfate groups are covalently bonded, the total charge per particle is constant and the charge density increases upon shrinking \cite{PeltonLangmuir1989, DalyPCCP2000}. For IPN microgels a similar mechanism still holds for the temperature behaviour of the electrophoretic mobility. Moreover, due to additional charged groups, belonging to AAc
moieties,  more negative values are found for C$_{PAAc}$ $\geq$ 2.6 \%. 
\begin{table}[b]
\small
  \begin{center}
     \begin{tabular}{c || c c c c c}
       \hline
	      {} & {$A_1$} & {$A_2$} & {$T_0$} & {$dT$} \\ 
				\hline
				\hline
        PNIPAM & -0.55 $\pm$ 0.03 & -2.8 $\pm$ 0.04 & 308.8 $\pm$ 0.2 & 1.7 $\pm$ 0.14  \\
        IPN 2.6\%  & -0.39 $\pm$ 0.04  & -2.9 $\pm$ 0.04 & 308.7 $\pm$ 0.1 & 1.2 $\pm$ 0.1  \\
        IPN 10.6\%  & -1.3 $\pm$ 0.04  & -3.8 $\pm$ 0.04 & 308.3 $\pm$ 0.1 & 1.6 $\pm$ 0.2 \\
        IPN 19.2\%  & -1.1 $\pm$ 0.05  & -4.5 $\pm$ 0.07 & 307.8 $\pm$ 0.1 & 2.9 $\pm$ 0.2  \\ 
       \hline
	
     \end{tabular}
     \caption{Values of the fitting parameters as obtained from the best fit of the temperature behaviour of the electrophoretic mobility through the sigmoidal function reported in the text: $y = A_2 + (A_1-A_2)/(1 + exp((T-T_0)/dT))$.} \label{tab:MobFit}   \end{center}
	\end{table}
We note that the electrokinetic transition temperature T$_0$ is higher than the VPTT in agreement with previous results on PNIPAM-based microgels \cite{PeltonLangmuir1989, DalyPCCP2000, HoarePolym2005}, showing a shift forward. This difference can be explained considering that the effective charge carriers are mainly confined to the peripheral shell and that the charged and less dense shell fully collapses at the end of the VPT (see Fig. \ref{fig:mobility}(b)).
This picture well describes our results for both PNIPAM and IPN microgels where a highly dense core of interpenetrated PNIPAM and PAAc networks is surrounded by a less dense shell mainly composed by PAAc chains.
Interestingly, the magnitude of the variation of mobility is dependent on PAAc content and it is more pronounced at high PAAc, where collapsed IPN microgels are characterized by more negative mobility values, as expected from the increase of the charge density due to the greater fraction of exposed PAAc chains above the VPTT. These results further support the possibility to control the effective charge density on the microgel surfaces through the amount of poly(acrylic acid) interpenetrating the PNIPAM network and hence also inter-particle interactions and aggregation phenomena.

\subsection*{Dependence on PAAc content}

\begin{figure}[!t]
\centering
\includegraphics[height=10cm]{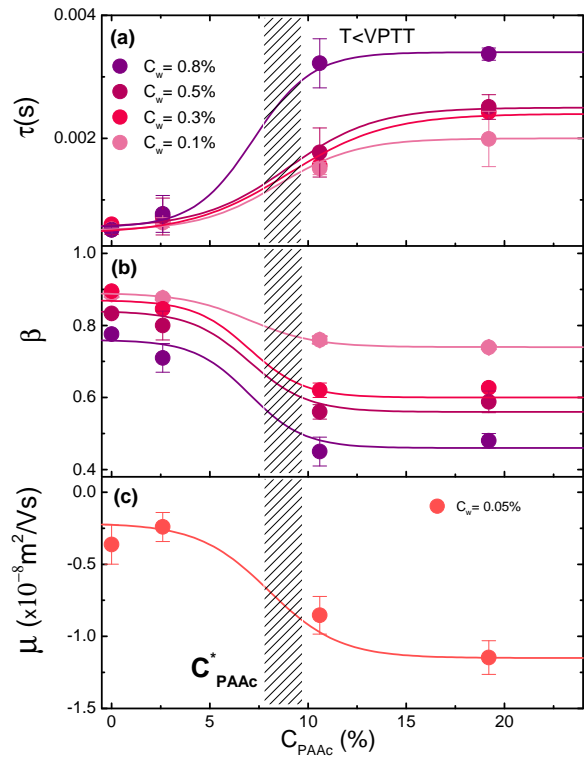}
\caption{(a) Relaxation time, (b) stretching parameter and (c) electrophoretic mobility as a function of the PAAc content at the indicated weight concentrations at pH 5.5 below the VPTT (T=295 K). Solid lines are guides to eyes.}
\label{fig:tau_beta_mob}
\end{figure}

To highlight the dynamical changes related to the interpenetration of the poly(acrylic acid) within the PNIPAM network, we report the behaviour of the relaxation time $\tau$, of the stretching parameter $\beta$  and of the electrophoretic mobility $\mu$ as a function of PAAc content and at fixed temperature below the VPTT in Fig.\ref{fig:tau_beta_mob}. Data exhibit a dramatic jump in a range of PAAc contents between $C_{PAAc}$=2.6 \% and $C_{PAAc}$=10.6 \% evidencing a cross-over between two different regions. A sigmoidal fit of the data suggests that a critical value of PAAc content has to be expected around $C_{PAAc}^* \approx$ 8 \% \footnote{$C_{PAAc}^*$: crossover PAAc content}: below this $C_{PAAc}^*$  IPN microgels behave very similarly to pure PNIPAM microgel, indicating that the charges influence is negligible, while above $C_{PAAc}^*$ the effect of PAAc, and therefore of charge density, becomes relevant leading to a slowing down of the dynamics (increase of the relaxation time), an enhancement of polydispersity (decrease of the stretching parameter of Eq.(\ref{Eqfit})) and an increase of the overall charge density (more negative values of mobility).

\begin{figure}[!t]
\centering
\includegraphics[height=7cm]{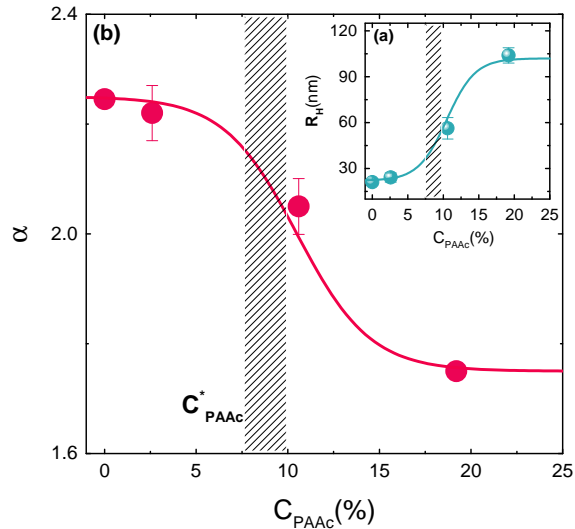}
\caption{Swelling ratio for IPN microgel as a function of the PAAc content at $C_w$=0.1 \% and pH 5.5. Inset: hydrodynamic radius R$_H$ in the shrunken state from DLS measurements as a function of the PAAc content. Solid lines are guides to eyes.}
\label{fig:RvsPAAc}
\end{figure}

A similar transition is observed in the microgel hydrodynamic radius R$_H$ \footnote{R$_H$: hydrodynamic radius} in the shrunken state shown in Fig.\ref{fig:RvsPAAc}(a).
The increase of R$_H$ with PAAc content can be explained considering that the structure of IPN microgels is characterized by a highly dense core of interpenetrated PNIPAM and PAAc networks surrounded by a low density shell mainly populated by PAAc dangling chains that increases in size as PAAc content increases.
The trend is also confirmed by the PAAc content dependence of the swelling ratio defined as:

\begin{equation}
\alpha = \frac{R_H^{swollen}}{R_H^{shrunken}}
\end{equation}

where $R_H^{swollen}=R(297 K)$ \footnote{$R_H^{swollen}$: hydrodynamic radius in the swollen state} and $R_H^{shrunken}=R(311 K)$\footnote{$R_H^{shrunken}$: hydrodynamic radius in the shrunken state}.
In  Fig.\ref{fig:RvsPAAc}(b) $\alpha$ \footnote{$\alpha$: swelling ratio} is reported at different PAAc contents, a clear decrease with increasing $C_{PAAc}$ is observed, indicating a reduction of the swelling capability of IPN microgels by increasing the PAAc amount interpenetrating PNIPAM network and thus leading to a decreases of particle softness with increasing topological constraints and charges due to the interpenetration of the two networks.

\section*{Conclusions}
\label{Conclusions}

At variance with previous studies \cite{XiaLangmuir2004, HuAdvMater2004, XiaJCRel2005, ZhouBio2008,MattssonNature2009,LiuPolymers2012, NigroJNCS2015, NigroJCP2015, NigroSM2017, NigroCSA2017} this work has demonstrated, through dynamic light scattering, transmission electron microscopy and electrophoretic measurements, the role played  by poly(acrylic acid) on the relaxation dynamics and on the aggregation processes in IPN microgels across the VPT. 
A crossover PAAc concentration $C_{PAAc}^*$ that signs the existence of two different regions, has been identified: below $C_{PAAc}^*$ IPN microgels behave very similarly to pure PNIPAM, while above $C_{PAAc}^*$ they significantly differ, indicating a stronger influence of the extent of ionic charges on the microgels. In fact, by increasing PAAc content, the effective charge density increases, as shown by electrophoretic mobility measurements and attractive interactions between protonated COOH and deprotonated COO$^-$ groups belonging to different particles are enhanced. This is reflected in the huge growth of the relaxation time signature of an aggregation process. The influence of the extent of ionic charges is also confirmed by the comparison between experimental data and theoretical models from the Flory-Rehner theory that shows as the number of counterions per polymer chain, f, increases with PAAc. Finally we prove that particle softness can be experimentally controlled by changing the PAAc/PNIPAM ratio: particles with low PAAc content are more soft and deformable while stiffer particles can be obtained by increasing the amount of PAAc chains interpenetrating PNIPAM network. Further studies with salt are under way to investigate if adding extra charges in the system through a monovalent salt has the same effect of increasing PAAc content and if the aggregation process is speeded up or slowed down.

The importance of our findings is twofold. On one hand our results are relevant for fundamental studies aiming to understand the role of softness in exotic phase diagrams with respect to hard colloids. On the other hand, the accurate knowledge of aggregation conditions with respect to parameters such as particle and polymeric concentration, pH and temperature is crucial to improve the manifold technological applications of these stimuli-responsive materials. In particular our work meets the request of Ref.\cite{VinogradovCurrPharmDes2006} to develop new nanogels as intelligent drug carriers. It will be very interesting to assess the ability of our multi responsive IPN microgel to encapsulate drugs and other small molecules, to deliver them and to investigate the effect of softness and charge (peculiar of these microgels) on their release.

\section*{Acknowledgments}
The authors acknowledge support from MIUR Fare SOFTART (R16XLE2X3L)

\section*{Author contributions statement}

R.A., V.N. and B.Ru. conceived the experiments. S.C., V.N., B.Ro. and S.S. conducted the experiments and analysed the results. M.B. and E.B. synthesized the samples. R.A., V.N., B.Ru. and S.S. reviewed the manuscript.

\section*{Additional information}

The authors declare no competing interests.

\subsection*{Bibliography}
\bibliographystyle{unsrt}\biboptions{sort&compress}

\begin{thebibliography}{10}

\bibitem{VinogradovCurrPharmDes2006}
S.~V. Vinogradov.
\newblock {Colloidal microgels in drug delivery applications}.
\newblock {\em Curr. Pharm. Des.}, 12:4703\textendash{}4712, 2006.

\bibitem{DasAnnRevMR2006}
M.~Das, H.~Zhang, and E.~Kumacheva.
\newblock {MICROGELS: Old Materials with New Applications}.
\newblock {\em Annu. Rev. Mater. Res.}, 36:117\textendash{}142, 2006.

\bibitem{ParkBiomat2013}
J.~S. Park, H.~N. Yang, D.~G. Woo, S.~Y. Jeon, and K.~H. Park.
\newblock {Poly(N-isopropylacrylamide-co-acrylic acid) nanogels for tracing and
  delivering genes to human mesenchymal stem cells}.
\newblock {\em Biomaterials}, 34:8819\textendash{}8834, 2013.

\bibitem{HamidiDrugDeliv2008}
M.~Hamidi, A.~Azadi, and P.~Rafie.
\newblock {Hydrogel nanoparticles in drug delivery}.
\newblock {\em Adv. Drug Deliv. Rev.}, 60:1638\textendash{}1649, 2008.

\bibitem{SmeetsPolymSci2013}
N.~M.~B. Smeets and T.~Hoare.
\newblock {Designing Responsive Microgels for Drug Delivery Applications}.
\newblock {\em J. Polym. Sci. A Polym. Chem.}, 51:3027\textendash{}3043, 2013.

\bibitem{SuBiomacro2008}
S.~Su, Ali~Md. Monsur, C.~D.~M. Filipe, Y.~Li, and R.~H. Pelton.
\newblock {Microgel-Based Inks for Paper-Supported Biosensing Applications}.
\newblock {\em Biomacromolecules}, 9:935\textendash{}941, 2008.

\bibitem{LikosJPCM2002}
C.~N. Likos, N.~Hoffmann, H.~L{\"{o}}wen, and A.~A. Louis.
\newblock {Exotic fluids and crystals of soft polymeric colloids}.
\newblock {\em J. Phys. Cond. Matter}, 14:7681--7698, 2002.

\bibitem{RamirezJPCM2009}
P.~E. Ram\'irez-Gonz\'alez and M.~Medina-Noyola.
\newblock Glass transition in soft-sphere dispersions.
\newblock {\em Journal of Physics: Condensed Matter}, 21(7):075101--13, 2009.

\bibitem{HeyesSM2009}
D.M. Heyes and A.C.Branka.
\newblock {Interactions between microgel particles}.
\newblock {\em Soft Matter}, 5:2681, 2009.

\bibitem{WangChemPhys2014}
H.~Wang, X.~Wu, Z.~Zhu, C.~S. Liu, and Z.~Zhang.
\newblock {Revisit to phase diagram of poly(N-isopropylacrylamide) microgel
  suspensions by mechanical spectroscopy}.
\newblock {\em J. Chem. Phys.}, 140:024908, 2014.

\bibitem{PritiJCP2014}
P.~S. Mohanty, D.~Paloli, J.~J. Crassous, E.~Zaccarelli, and P.~Schurtenberger.
\newblock {Effective interactions between soft-repulsive colloids: Experiments,
  theory and simulations}.
\newblock {\em J. Chem. Phys.}, 140:094901, 2014.

\bibitem{HellwegCPS2000}
T.~Hellweg, C.D. Dewhurst, E.~Br{\"u}ckner, K.Kratz, and W.Eimer.
\newblock {Colloidal crystals made of poly(N-isopropylacrylamide) microgel
  particles}.
\newblock {\em Colloid. Polym. Sci.}, 278:972--978, 2000.

\bibitem{PaloliSM2013}
D.~Paloli, P.~S. Mohanty, J.~J. Crassous, E.~Zaccarelli, and P.~Schurtenberger.
\newblock {Fluid\textendash{}solid transitions in soft-repulsive colloids}.
\newblock {\em Soft Matter}, 9:3000--3004, 2013.

\bibitem{LyonRevPC2012}
L.~A. Lyon and A.~Fernandez-Nieves.
\newblock {The Polymer/Colloid Duality of Microgel Suspensions}.
\newblock {\em Annu. Rev. Phys. Chem.}, 63:25--43, 2012.

\bibitem{WuPRL2003}
J.~Wu, B.~Zhou, and Z.~Hu.
\newblock {Phase behavior of thermally responsive microgel colloids}.
\newblock {\em Phys. Rev. Lett.}, 90(4):048304--4, 2003.

\bibitem{PuseyNat1986}
P.~N. Pusey and W.~van Megen.
\newblock Phase behaviour of concentrated suspensions of nearly hard colloidal
  spheres.
\newblock {\em Nature}, 320:340--342, 1986.

\bibitem{ImhofPRL1995}
A.~Imhof and J.~K.~G. Dhont.
\newblock {Experimental Phase Diagram of a Binary Colloidal Hard-Sphere Mixture
  with a Large Size Ratio}.
\newblock {\em Phys. Rev. Lett.}, 75:1662--1665, 1995.

\bibitem{PhamScience2002}
K.~N. Pham, A.~M. Puertas, J.~Bergenholtz, S.~U. Egelhaaf, A.~Moussa{\"i}d,
  P.~N. Pusey, A.~B. Schofield, M.~E. Cates, M.~Fuchs, and W.~C.~K. Poon.
\newblock {Multiple Glassy States in a Simple Model System}.
\newblock {\em Science}, 296:104--106, 2002.

\bibitem{EckertPRL2002}
T.~Eckert and E.~Bartsch.
\newblock {Re-entrant glass transition in a colloid-polymer mixture with
  depletion attractions}.
\newblock {\em Phys. Rev. Lett.}, 89:125701--4, 2002.

\bibitem{LuNat2008}
P.~J. Lu, E.~Zaccarelli, F.~Ciulla, A.~B. Schofield, F.~Sciortino, and D.~A.
  Weitz.
\newblock Gelation of particle with short range attraction.
\newblock {\em Nature}, 453:499--503, 2008.

\bibitem{RoyallNatMat2008}
C.~P. Royall, S.~R. Williams, T.~Ohtsuka, and H.~Tanaka.
\newblock {Direct observation of a local structural mechanism for dynamical
  arrest}.
\newblock {\em Nat. Mater.}, 7:556--561, 2008.

\bibitem{RuzickaNatMat2011}
B.~Ruzicka, E.~Zaccarelli, L.~Zulian, R.~Angelini, M.~Sztucki, A.~Moussa{\"i}d,
  T.~Narayanan, and F.~Sciortino.
\newblock {Observation of empty liquids and equilibrium gels in a colloidal
  clay}.
\newblock {\em Nat. Mater.}, 10:56--60, 2011.

\bibitem{AngeliniNC2014}
R.~Angelini, E.~Zaccarelli, F.~A. de~Melo~Marques, M.~Sztucki, A.~Fluerasu,
  G.~Ruocco, and B.~Ruzicka.
\newblock {Glass-glass transition during aging of a colloidal clay}.
\newblock {\em Nat. Commun.}, 5:4049--7, 2014.

\bibitem{BanchioJCP2008}
{A. J. Banchio and G. N\"{a}gele}.
\newblock {Short-time transport properties in dense suspensions: From neutral
  to charge-stabilized colloidal spheres}.
\newblock {\em {J. Chem. Phys}}, {128}:{104903}, {2008}.

\bibitem{GapinskiJCP2009}
{J. Gapinski and A. Patkowski and A. J. Banchio and J. Buitenhuis and P.
  Holmqvist and M. P. Lettinga and G. Meier and G. N\"{a}gele}.
\newblock {Structure and short-time dynamics in suspensions of charged silica
  spheres in the entire fluid regime}.
\newblock {\em J. Chem. Phys}, 130:084503, 2009.

\bibitem{MaColloidInt2010}
J.~Ma, B.~Fan, B.~Liang, and J.~Xu.
\newblock {Synthesis and characterization of
  Poly(N-isopropylacrylamide)/Poly(acrylic acid) semi-IPN nanocomposite
  microgels}.
\newblock {\em J. Colloid Interface Sci.}, 341:88\textendash{}93, 2010.

\bibitem{ShibayamaJCP1992}
M.~Shibayama, T.~Tanaka, and C.~C. Han.
\newblock {Small angle neutron scattering study of poly(N-isopropyl acrylamide)
  gels near their volume-phase transition temperature}.
\newblock {\em J. Chem. Phys}, 97:6829\textendash{}6841, 1992.

\bibitem{KratzBerBunsenges11998}
K.~Kratz and W.~Eimer.
\newblock {Swelling Properties of Colloidal Poly(N-Isopropylacrylamide)
  Microgels in Solution}.
\newblock {\em Ber. Bunsenges. Phys. Chem.}, 102:848\textendash{}854, 1998.

\bibitem{KratzPolymer2001}
K.~Kratz, T.~Hellweg, and W.~Eimer.
\newblock {Structural changes in PNIPAM microgel particles as seen by SANS, DLS
  and EM techniques}.
\newblock {\em Polymer}, 42:6631\textendash{}6639, 2001.

\bibitem{KratzColloids2000}
K.~Kratz, T.~Hellweg, and W.~Eimer.
\newblock {Influence of charge density on the swelling of colloidal
  poly(N-isopropylacrylamide-co-acrylic acid) microgels}.
\newblock {\em Colloids Surf. A}, 170:137\textendash{}149, 2000.

\bibitem{KratzBerBunsenges21998}
K.~Kratz, T.~Hellweg, and W.~Eimer.
\newblock {Effect of connectivity and charge density on the swelling and local
  structure and dynamic properties of colloidal PNIPAM microgels}.
\newblock {\em Ber. Bunsenges. Phys. Chem.}, 102(11):1603\textendash{}1608,
  1998.

\bibitem{JonesMacromol2000}
C.~D. Jones and L.~A. Lyon.
\newblock {Synthesis and Characterization of Multiresponsive Core-Shell
  Microgels}.
\newblock {\em Macromolecules}, 33:8301\textendash{}8303, 2000.

\bibitem{XiongColloidSurf2011}
W.~Xiong, X.~Gao, Y.~Zao, H.~Xu, and X.~Yang.
\newblock {The dual temperature/pH-sensitive multiphase behavior of
  poly(Nisopropylacrylamide-co-acrylic acid) microgels for potential
  application in \textit{in situ} gelling system}.
\newblock {\em Colloids Surf. B: Biointerfaces}, 84:103\textendash{}110, 2011.

\bibitem{MengPhysChem2007}
Z.~Meng, J.~K. Cho, S.~Debord, V.~Breedveld, and L.~A. Lyon.
\newblock {Crystallization Behavior of Soft, Attractive Microgels}.
\newblock {\em J. Phys. Chem. B}, 111:6992\textendash{}6997, 2007.

\bibitem{LyonJPCB2004}
L.~A. Lyon, J.~D. Debord, S.~B. Debord, C.~D. Jones, J.~G. McGrath, and M.~J.
  Serpe.
\newblock {Microgel Colloidal Crystals}.
\newblock {\em J. Phys. Chem. B}, 108:19099\textendash{}19108, 2004.

\bibitem{HolmqvistPRL2012}
P.~Holmqvist, P.~S. Mohanty, G.~N{\"a}gele, P.~Schurtenberger, and M.~Heinen.
\newblock {Structure and Dynamics of Loosely Cross-Linked Ionic Microgel
  Dispersions in the Fluid Regime}.
\newblock {\em Phys. Rev. Lett.}, 109:048302--5, 2012.

\bibitem{DebordJPCB2003}
S.~B. Debord and L.~A. Lyon.
\newblock {Influence of Particle Volume Fraction on Packing in Responsive
  Hydrogel Colloidal Crystals}.
\newblock {\em J. Phys. Chem. B}, 107:2927\textendash{}2932, 2003.

\bibitem{HuAdvMater2004}
Z.~Hu and X.~Xia.
\newblock {Hydrogel nanoparticle dispersions with inverse thermoreversible
  gelation}.
\newblock {\em Adv. Mater.}, 16(4):305\textendash{}309, 2004.

\bibitem{XiaLangmuir2004}
X.~Xia and Z.~Hu.
\newblock {Synthesis and Light Scattering Study of Microgels with
  Interpenetrating Polymer Networks}.
\newblock {\em Langmuir}, 20:2094\textendash{}2098, 2004.

\bibitem{XiaJCRel2005}
X.~Xia, Z.~Hu, and M.~Marquez.
\newblock {Physically bonded nanoparticle networks: a novel drug delivery
  system}.
\newblock {\em J. Control. Release}, 103:21\textendash{}30, 2005.

\bibitem{ZhouBio2008}
J.~Zhou, G.~Wang, L.~Zou, L.~Tang, M.~Marquez, and Z.~Hu.
\newblock {Viscoelastic Behavior and In Vivo Release Study of Microgel
  Dispersions with Inverse Thermoreversible Gelation}.
\newblock {\em Biomacromolecules}, 9:142\textendash{}148, 2008.

\bibitem{XingCollPolym2010}
Z.~Xing, C.~Wang, J.~Yan, L.~Zhang, L.~Li, and L.~Zha.
\newblock {pH/temperature dual stimuli-responsive microcapsules with
  interpenetrating polymer network structure}.
\newblock {\em Colloid Polym. Sci.}, 288:1723\textendash{}1729, 2010.

\bibitem{LiuPolymers2012}
X.~Liu, H.~Guo, and L.~Zha.
\newblock {Study of pH/temperature dual stimuli-responsive nanogels with
  interpenetrating polymer network structure}.
\newblock {\em Polymers}, 61(7):1144\textendash{}1150, 2012.

\bibitem{NigroJNCS2015}
V.~Nigro, R.~Angelini, M.~Bertoldo, V.~Castelvetro, G.~Ruocco, and B.~Ruzicka.
\newblock {Dynamic light scattering study of temperature and pH sensitive
  colloidal microgels}.
\newblock {\em J. Non-Cryst. Solids}, 407:361--366, 2015.

\bibitem{NigroJCP2015}
V.~Nigro, R.~Angelini, M.~Bertoldo, F.~Bruni, M.A. Ricci, and B.~Ruzicka.
\newblock Local structure of temperature and ph-sensitive colloidal microgels.
\newblock {\em J. Chem. Phys.}, 143:114904--9, 2015.

\bibitem{NigroCSA2017}
V.~Nigro, R.~Angelini, M.~Bertoldo, and B.~Ruzicka.
\newblock Swelling of responsive-microgels: experiments versus models.
\newblock {\em Colloids Surf. A}, 532:389--396, 2017.

\bibitem{NigroSM2017}
V.~Nigro, R.~Angelini, M.~Bertoldo, F.~Bruni, M.A. Ricci, and B.~Ruzicka.
\newblock Dynamical behavior of microgels of interpenetrated polymer networks.
\newblock {\em Soft Matter}, 13:5185--5193, 2017.

\bibitem{MattssonNature2009}
J.~Mattsson, H.~M. Wyss, A.~Fernandez-Nieves, K.~Miyazaki, Z.~Hu, D.~Reichman,
  and D.~A. Weitz.
\newblock {Soft colloids make strong glasses}.
\newblock {\em Nature}, 462(5):83\textendash{}86, 2009.

\bibitem{VillariCPC2018}
N.~Micali, M.~Bertoldo, E.~Buratti, V.~Nigro, R.~Angelini, and V.~Villari.
\newblock Interpenetrating polymer network microgels in water: effect of
  composition on the structural properties and electrosteric interactions.
\newblock {\em ChemPhysChem}, 19:2894--2901, 2018.

\bibitem{NotaCampioni}
In previous works the same samples were reported with the following PAAc concentrations: $C_{PAAc}=4 \%$, $C_{PAAc}=9 \%$, $C_{PAAc}=23 \%$. These values were estimated before performing the more accurate elemental analysis.

\bibitem{AnderssonJPolymSci2006}
M.~Andersson and S.~L. Maunu.
\newblock {Structural Studies of Poly(N-isopropylacrylamide) Microgels: Effect
  of SDS Surfactant Concentration in the Microgel Synthesis.}
\newblock {\em J. Polym. Sci. Part B Polym. Phys.}, 44:3305--3314, 2006.

\bibitem{KohlrauschAnnPhys1854}
R.~Kohlrausch.
\newblock {Theorie des elektrischen rckstandes in der leidener flasche}.
\newblock {\em Annalen der Physik,}, 2:179--214, 1854.

\bibitem{WilliamsFaradayTrans1970}
G.~Williams and D.~C. Watts.
\newblock {Non-Symmetrical Dielectric Relaxation Behavior Arising from a Simple
  Empirical Decay Function}.
\newblock {\em J. Chem. Soc. Faraday Trans.}, 66:80\textendash{}85, 1970.

\bibitem{TscharnuterAppOpt2001}
W.W. Tscharnuter.
\newblock Mobility measurements by phase analysis.
\newblock {\em Applied optics}, 40:3995--4003, 2001.

\bibitem{MinorJCIS1997}
M.~Minor, A.J. van~der Linde, H.P. van Leeuwen, and J.~Lyklema.
\newblock Dynamic aspects of electrophoresis and electroosmosis: A new fast
  method for measuring particle mobilities.
\newblock {\em J. Colloid Interface Sci.}, 189:370--375, 1997.

\bibitem{ConnahJDST2002}
M.~Connah, M.~Kaszuba, and A.~Morfesis.
\newblock High resolution of zeta potential measurements: Analysis of
  multi-component mixtures.
\newblock {\em J. Dispersion Sci. Technol.}, 23:663--669, 2002.

\bibitem{LopezSM2017}
{C.G. L{\`{o}}pez-Le{\`{o}}n and W. Richtering}.
\newblock {Does Flory-Rehner theory quantitatively describe the swelling of
  thermoresponsive microgels?}
\newblock {\em {Soft Matter}}, { 13}:{8271--8280 }, { 2017}.

\bibitem{Flory1953}
P.J. Flory.
\newblock {\em Principles of Polymer Chemistry}.
\newblock Cornell University, Ithaca, New York, 1953.

\bibitem{FernandezBook2011}
A.~Fernandez-Nieves, H.~Wyss, J.~Mattsson, and D.A. Weitz.
\newblock {\em Microgel Suspensions: Fundamentals and Applications}.
\newblock Wiley-VCH Verlag, 2011.

\bibitem{QuesadaPerezSM2011}
Manuel Quesada-P\'{e}rez, José~Alberto Maroto-Centeno, Jacqueline Forcada, and
  Roque Hidalgo-Alvarez.
\newblock Gel swelling theories: the classical formalism and recent approaches.
\newblock {\em Soft Matter}, 7:10536--10547, 2011.

\bibitem{LopezLeonPRE2007}
T.~L\`{o}pez-Le\`{o}n and A.~Fernandez-Nieves.
\newblock Macroscopically probing the entropic influence of ions: deswelling
  neutral microgels with salt.
\newblock {\em Phys. Rev. E}, 75:011801, 2007.

\bibitem{ErmanMacromol1986}
B.~Erman and P.J. Flory.
\newblock Critical phenomena and transitions in swollen polymer networks and in
  linear macromolecules.
\newblock {\em Macromolecules}, 19:2342--2353, 1986.

\bibitem{PeltonAdvColloid2000}
R.~H. Pelton.
\newblock {Temperature-sensitive aqueous microgels}.
\newblock {\em Adv. Colloid Interface Sci.}, 85:1\textendash{}33, 2000.

\bibitem{StiegerJCP2004}
M.~Stieger, W.~Richtering, J.S. Pedersen, and P.~Lindner.
\newblock Small-angle neutron scattering study of structural changes in
  temperature sensitive microgel colloids.
\newblock {\em The Journal of chemical physics}, 120(13):6197--6206, 2004.

\bibitem{MasonPRE2005}
T.~G. Mason, , and M.~Y. Lin.
\newblock Density profiles of temperature-sensitive microgel particles.
\newblock {\em Phys. Rev. E}, 71:040801, 2005.

\bibitem{ReuferEPJ2009}
M.~Reufer, P.~D{\i}az-Leyva, I.~Lynch, and F.~Scheffold.
\newblock Temperature-sensitive poly (n-isopropyl-acrylamide) microgel
  particles: A light scattering study.
\newblock {\em The European Physical Journal E: Soft Matter and Biological
  Physics}, 28(2):165--171, 2009.

\bibitem{LedesmaCSA2015}
M.~Ledesma-Motolin{\'\i}a, M.~Braibanti, L.F. Rojas-Ochoa, and
  C.~Haro-P{\'e}rez.
\newblock Interplay between internal structure and optical properties of
  thermosensitive nanogels.
\newblock {\em Colloids and Surfaces A: Physicochemical and Engineering
  Aspects}, 482:724--727, 2015.

\bibitem{RomeoSM2013}
G.~Romeo and M.~Pica Ciamarra.
\newblock {Elasticity of compressed microgel suspensions}.
\newblock {\em Soft Matter}, 9:5401--5406, 2013.

\bibitem{GrohnMacromol2000}
{ F. Gr{\"{o}}hn and M. Antonietti}.
\newblock {Intermolecular Structure of Spherical Polyelectrolyte Microgels in
  Salt-Free Solution. 1. Quantification of the Attraction between Equally
  Charged Polyelectrolytes}.
\newblock {\em { Macromolecules}}, {33}:{5938--5949 \color{black}}, {2000}.

\bibitem{PeltonLangmuir1989}
R.~H. Pelton, H.~M. Pelton, A.~Morphesis, and R.~L. Rowells.
\newblock Particle sizes and electrophoretic mobilities of poly(
  n-isopropylacrylamide) latex.
\newblock {\em Langmuir}, 5:816--818, 1989.

\bibitem{DalyPCCP2000}
E.~Daly and B.R. Saunders.
\newblock Temperature-dependent electrophoretic mobility and hydrodynamic
  radius measurements of poly(n-isopropylacrylamide) microgel particles:
  structural insights.
\newblock {\em Phys. Chem. Chem. Phys.}, 2:3187--3193, 2000.

\bibitem{HoarePolym2005}
T.~Hoare and R.~Pelton.
\newblock {Electrophoresis of functionalized microgels: morphological
  insights}.
\newblock {\em Polymer}, 46:1139--1150, 2005.

\end{thebibliography}

\end{document}